\documentclass[abstract=true, DIV=14]{scrartcl}

\usepackage{amsmath}
\usepackage{amssymb}
\usepackage{amsthm}
\usepackage{mathtools}
\usepackage[mathscr]{euscript}
\usepackage[shortlabels]{enumitem}
\usepackage{graphicx}

\usepackage{xcolor}
\usepackage{microtype}
\usepackage{algorithm}
\usepackage[noend]{algpseudocode}
\algrenewcommand\algorithmicrequire{\textbf{Input:}}
\algrenewcommand\algorithmicensure{\textbf{Output:}}

\usepackage{thm-restate}

\usepackage[T1]{fontenc}
\usepackage{lmodern}
\usepackage[utf8]{inputenc}

\usepackage[
    backend=biber,
    style=alphabetic,
    sorting=nyt,
    maxnames=99,
    maxcitenames=99,
    maxalphanames=4,
    minalphanames=3,
    url=false
]{biblatex}

\DeclareFieldFormat[article,inbook,incollection,inproceedings,patent,thesis,unpublished,misc]{title}{#1}
\DeclareFieldFormat[article,inbook,incollection,inproceedings]{pages}{pages #1}

\usepackage[normalem]{ulem}

\usepackage{upgreek}
\usepackage[colorlinks]{hyperref}
\usepackage[capitalize]{cleveref}

\newtheorem{theorem}{Theorem}[section]
\newtheorem{lemma}[theorem]{Lemma}

\crefname{equation}{eq.}{eqs.} 
\crefname{enumi}{}{} 

\crefname{icase}{case}{cases}
\crefname{ipart}{part}{parts}
\crefname{iprop}{property}{properties}
\crefname{iinv}{invariant}{invariants}

\crefformat{section}{\S\,#2#1#3}
\crefformat{subsection}{\S\,#2#1#3}
\crefrangeformat{section}{\S\S\,#3#1#4--#5#2#6}
\crefmultiformat{section}{\S\S\,#2#1#3}{,~#2#1#3}{,~#2#1#3}{,~#2#1#3}

\usepackage{authblk}

\newcommand{\fO}{\mathcal{O}}

\newcommand{\cP}{\mathcal{P}}

\newcommand{\eat}[1]{}

\DeclareMathOperator{\wid}{width}

\DeclareMathOperator{\first}{first} 
\DeclareMathOperator{\last}{last}

\renewcommand{\alpha}{\upalpha}

\newcommand{\cC}{\mathscr{C}}

\newcommand{\connected}[1]{\def\temp{#1}\ifx\temp\empty\sim\else\overset{#1}{\sim}\fi}

\title{Faster polynomial-space algorithms for poset problems}

\author[1]{L\'{a}szl\'{o} Kozma}
\author[2]{Junqi Tan}
\affil[1]{Dresden University of Technology}
\affil[2]{Freie Universit\"at Berlin}

\date{}

\begin{document}

\maketitle

\begin{abstract}
We give a polynomial-space, deterministic, exact algorithm for \emph{counting linear extensions} of partial orders over $n$ elements, with running time $\fO^*((2\sqrt{2})^{n}) \subseteq \fO(2.829^n)$. This improves upon the long-standing bound of $\fO^*(4^{n})$ attainable by standard divide and conquer (Savitch 1970, Gurevich and Shelah 1987). Our algorithm uses a similar \emph{balanced splitting} approach as the classical algorithm, but applies it only to the elements outside a largest antichain. For the elements of the antichain itself we develop a formula based on inclusion-exclusion that accounts for the ways in which they can appear in a linear extension. The technique is also applicable to other problems on posets, which we illustrate by solving the well-studied \emph{jump number} problem in a similar running time of $\fO^*((2\sqrt{2})^{n})$ and polynomial space. To our knowledge, these are the first polynomial-space algorithms with a running time below $4^n$ for both problems. 

\end{abstract}

\section{Introduction}\label{sec1}

A \emph{partially ordered set} (poset) $\mathcal{P}=(X,\prec)$ consists of a \emph{ground set} $X$ and a binary relation $\prec$ that describes precedences between elements of $X$. Posets are often used to model the intermediate state of comparison-based algorithms (such as for sorting), when some pairs have already been compared (or their ordering inferred), while the ordering of other pairs is still unknown. The relation $\prec$ is asymmetric ($a \prec b$ implies $b \not\prec a$) and transitive ($a \prec b$ and $b \prec c$ imply $a \prec c$).

A natural algorithmic problem is, given a poset $\cP$, to compute its number of linear extensions $e(\cP)$, i.e., the number of total orders of $X$ consistent with $\prec$. We refer to this problem as \#LE.

For general background on posets we refer to the monographs~\cite{Trotter_book, Rival_book}, \cite[\S\,3]{Stanley_book}, \cite[\S\,8]{Handbook_comb}.  

Counting linear extensions arises in combinatorics, e.g., in the context of the famous $1/3$--$2/3$-conjecture~\cite{Brightwell_survey, KahnSaks, Brightwell1995}, as well as in applications such as learning graphical models~\cite{Wallace, Niinimaki}, automated  planning~\cite{DBLP:journals/jair/MuiseBM16}, probabilistic ranking~\cite{Winkler_avg, Fishburn, Lukasiewicz}, order-reconstruction in data mining~\cite{MannilaMeek}, convex rank tests in non-parametric statistics~\cite{Morton}, multimedia delivery in networks~\cite{Amer}, and others. Counting linear extensions (alternatively: counting topological orderings of a directed acyclic graph) is a well-studied problem, both in exact and approximate settings (early references are Linial~\cite{Linial} and Lov\'asz~\cite[§\,2.4]{Lovasz}); an elegant polytope-formulation of \#LE is given by Stanley~\cite{Stanley_polytope}. The exact \#LE problem was shown to be \#P-complete by Brightwell and Winkler in 1991~\cite{Brightwell1991}. 

While the number of linear extensions of an  $n$-element poset can be as large as $n!$, \#LE can be solved faster than enumeration, in particular, in $\fO^*(2^n)$ time\footnote{The $\fO^*$-notation hides factors polynomial in $n$.} by the classical dynamic programming approach due to Bellman~\cite{Bellman1962}, and Held and Karp~\cite{HeldKarp1962}; see also~\cite{KSZ, DeLoof}. This involves iterating over all possible \emph{prefix-sets} of a linear order (ideals of $\cP$), counting possible ways of extending the prefix by a single element. The worst-case space requirement of the algorithm is also of the form $\fO^*(2^n)$, which is often an even greater impediment to practical implementation than the exponential runtime. 

An alternative approach based on divide and conquer is due to Savitch~\cite{savitch1970}, and Gurevich and Shelah~\cite{GurevichShelah1987}. It reduces the space requirement to polynomial, at the cost of an increased running time of $\fO^*(4^n)$. 
It involves iterating through all possible ways of splitting $X$ into two equal parts, where one part can entirely precede the other, according to $\prec$, and recursively solving the problem on both parts.

Both algorithms are applicable, with minimal adjustments, to a broad family of \emph{permutation-type problems}, most notably to the traveling salesman problem (TSP). In full generality, no improvements are known to these running times. In particular, solving TSP in $(2-\varepsilon)^n$ time or in $(4-\varepsilon)^n$ time and polynomial space are considered central open questions (e.g., see~\cite{FominKratsch, nederlof2026}). The two algorithms are at opposite extremes of a natural time-space tradeoff of $T\cdot S = 4$, between running time $T^n$ and space complexity $S^n$. Improved tradeoffs are known at intermediate points, achieving $T \cdot S < 4$ for a broad family of problems, including both TSP and \#LE~\cite{KoivistoParviainen2010, Ameli, Dallant}.  

For \#LE, Oka~\cite{Oka} recently  described an algorithm with running time and space requirement of the form $\fO(1.791^n)$, breaking the natural barrier of $2^n$. This result, based on dynamic programming, does not imply a polynomial-space algorithm, and a runtime improvement in this regime remained open. In this paper we give a new polynomial-space algorithm, significantly improving the classical $4^n$ bound.

\begin{theorem}\label{thm1}
The number of linear extensions of an $n$-element poset can be computed in time $\fO^*((2\sqrt{2})^n) \subseteq \fO(2.829^n)$ and polynomial space.
\end{theorem}

As its outer loop, our algorithm uses a balanced divide-and-conquer approach similar to the one described above, but splits only the complement of a maximum antichain. For the elements of the antichain itself, we compactly express the number of ways in which they can be inserted into an ordering of the remaining elements by an inclusion-exclusion formula. This costs a factor of at most $2^w$, where $w$ is the size of the antichain, i.e., the width of the poset, but the recursion then only needs to split the remaining $n-w$ elements. These elements can be partitioned into at most $w$ chains, which limits the number of balanced splits. Both small and large widths lead to savings, and our analysis captures the two effects at every level of the recursion, without an explicit case-distinction. 

The width also plays a role in the algorithm of Oka~\cite{Oka} and in earlier work on poset problems~\cite{poset2d, jump}. Their treatment of large antichains, however, is different, using techniques specific to dynamic programming that require an exponential number of states.

Our result holds for general posets; faster algorithms for \#LE are known for various special cases, including \emph{series-parallel}~\cite{Mohring}, \emph{low treewidth}~\cite{LE_tw, LE_sparse, LE_treewidth3}, \emph{small width}~\cite{DeLoof}, avoiding certain \emph{substructures}~\cite{LE_nfree}, and \emph{$2$-dimensional}~\cite{poset2d} posets. The latter result, in particular, is the starting point of our approach, as it already identified large width as the main bottleneck of \#LE, but addressed it only in a limited geometric setting.

\paragraph{The jump number problem.}
Let $\pi=(x_1,\dots,x_n)$ be a linear extension of $\mathcal{P}$. A pair $(x_i,x_{i+1})$ of consecutive elements is a \emph{bump} if $x_i\prec x_{i+1}$, and a \emph{jump} otherwise, i.e., if $x_i$ and $x_{i+1}$ are incomparable. The \emph{jump number} $\mathrm{jump}(\mathcal{P})$ is the minimum number of jumps in a linear extension of $\mathcal{P}$. As every linear extension has $n-1$ pairs of consecutive elements, minimizing the number of jumps is the same as maximizing the number of bumps.

The problem has its origins in scheduling~\cite{Pul81,CP85}: jobs with precedence constraints are processed one at a time on a single machine, and a job that does \emph{not} depend on the job processed just before it (i.e., a jump) requires a costlier setup. The jump number is also a well-studied invariant of posets~\cite{CH80,Mohring}; for instance, the jumps of a linear extension cut it into chains, so $\mathrm{jump}(\mathcal{P})\ge\mathrm{width}(\mathcal{P})-1$.

Pulleyblank~\cite{Pul81} showed that computing the jump number is NP-hard (see also~\cite{BH87}), and this holds even for interval orders~\cite{Mit91}. Polynomial-time algorithms are known for several classes of posets, among them \emph{N-free}~\cite{Riv83}, \emph{cycle-free}~\cite{DRW82}, and \emph{bounded width}~\cite{CP85} posets. The problem is fixed-parameter tractable when parameterized by the jump number~\cite{McC01} (see also~\cite{ES84}).

Similarly to \#LE, dynamic programming over the ideals of $\mathcal{P}$ computes the jump number in time and space $\mathcal{O}^*(2^n)$. Kratsch and Kratsch~\cite{jump} improved the running time to $\mathcal{O}(1.8638^n)$, and this was later improved to $\mathcal{O}(1.824^n)$~\cite{poset2d}. These algorithms use exponential space. The divide-and-conquer approach of Savitch~\cite{savitch1970} and Gurevich and Shelah~\cite{GurevichShelah1987} extends to the jump number, yielding polynomial space and running time $\mathcal{O}^*(4^n)$. We show the following.

\begin{theorem}\label{thm:jump}
The jump number of an $n$-element poset can be computed in time $\mathcal{O}^*((2\sqrt{2})^n)\subseteq\mathcal{O}(2.829^n)$ and polynomial space.
\end{theorem}

Within the same bounds, the algorithm also finds a linear extension with the minimum number of jumps.
Our algorithm for Theorem~\ref{thm:jump} (\S\,\ref{sec:jump}) follows the approach for \#LE: we remove a maximum antichain $B$ and split the orderings of the remaining elements in the middle, passing a small vector of partial results from one half to the other. The antichain elements are easier to handle than for \#LE. An element $x\in B$ can form a bump only right after the last of its predecessors or right before the first of its successors. So instead of counting the ways of inserting $B$, it suffices to handle these two positions while walking along an ordering of the remaining elements, and the delicate inclusion-exclusion step is not needed. 

\section{Preliminaries and algorithmic framework}\label{sec2}

Let $\cP = (X, \prec)$ be a poset and let $n = |X|$ denote its number of elements. A linear extension of $\cP$ is an ordering of $X$ in which $x$ precedes $y$ whenever $x \prec y$, for all $x,y \in X$. The number of linear extensions of $\cP$ is denoted $e(\cP)$. 

For $Y \subseteq X$, we denote by $\cP[Y]$ the subposet induced by $Y$. A set $I \subseteq X$ is an \emph{ideal} (or downset) if $y \in I$ and $x \prec y$ imply $x \in I$. Every prefix of a linear extension forms an ideal and every ideal is a prefix of some linear extension. 

An \emph{antichain} of $\cP$ is a subset of $X$ consisting of pairwise incomparable elements, and $\wid(\cP)$ denotes the size of the largest antichain. By Dilworth's theorem, $\cP$ can be decomposed into $\wid(\cP)$ chains. Both $\wid(\cP)$ and a corresponding chain partition of $\cP$ can be computed in polynomial time using bipartite matching~\cite{Schrijver2003Ch14}. 
At the two extremes, $\wid(\cP)=1$ when $\cP$ is a chain, and $\wid(\cP)=n$ when $\cP$ is an antichain.

\paragraph{Balanced splitting.}

The following simple formula forms the basis of the counting algorithm developed in this paper, similarly to the classical divide and conquer for permutation-type problems~\cite{savitch1970,GurevichShelah1987}.

\begin{lemma}\label{lem_a}
For every $m$ with $0 \leq m \leq n$,
$$e(\cP) = \sum_{\substack{I \mbox{~an ideal of~} \cP,\\|I| = m}}{e(\cP[I]) \cdot e(\cP[X \setminus I])}. $$
\end{lemma}
\begin{proof}
Every linear extension of $\cP$ splits into a prefix of $m$ elements that form a linear extension of $\cP[I]$ for some ideal $I$ and the remaining $n-m$ elements that form the linear extension of the remainder $\cP[X \setminus I]$. Conversely, for any ideal $I$, any linear extension of $\cP[I]$ followed by any linear extension of $\cP[X \setminus I]$ is a valid linear extension, since $y \prec x$ is not possible for any $x \in I$ and $y \in X \setminus I$.
\end{proof}

We obtain a baseline algorithm (essentially the classical approach) by applying Lemma~\ref{lem_a} recursively, as shown in Algorithm~\ref{alg_a}. 
As there are $\fO(\log{n})$ recursive levels with minimal bookkeeping in each call, the total space requirement is clearly polynomial. 

The running time $T(n)$ satisfies the inequality $T(n) \leq 2^n (2T(n/2) + \mathrm{poly}(n))$, resolving to $\fO^*(4^n)$. 

\begin{algorithm}[h]
\caption{Baseline divide and conquer algorithm}
\label{alg_a}
\begin{algorithmic}[1]
\Function{Split}{$\cP$}
    \State \textbf{if~}{$n \leq 1$}\textbf{~then}
        \State \quad \Return $1$
 
    \State $\mathit{count~}\gets 0$
    
    \State\textbf{for each}~ideal~$I$~of~$\cP$~of~size~$\lfloor n/2 \rfloor$
        \State \quad $\mathit{~count~}\gets \mathit{~count~} +~ \Call{Split}{\cP[I]} \cdot \Call{Split}{\cP[X \setminus I]}$

    \State \Return $\mathit{count}$
\EndFunction
\end{algorithmic}
\end{algorithm}

Our improved algorithm excludes a maximum antichain $B$ from the recursion. Then, similarly to Algorithm~\ref{alg_a}, it splits the remaining elements into two roughly equal parts, one of which forms an ideal, and recursively counts the orderings of the two parts. Finally, it counts the number of ways in which the elements of $B$ can be inserted into the combined orderings of the remaining elements. For the latter task we develop an inclusion-exclusion-type formula. 

If $B$ is large, then its complement is small and the recursion is cheap. If $B$ is small, then the remaining elements can be decomposed into a small number of chains, so they have few ideals, and the balanced splitting itself is efficient. We describe the algorithm in \S\,\ref{sec4} and analyze its running time in \S\,\ref{sec5}. In \S\,\ref{sec:jump} we adapt the approach to the jump number problem.

\section{Excluding a maximum antichain}\label{sec4}

When the poset $\cP$ has a large antichain, it can have many half-size ideals, so the balanced splitting approach alone is inefficient. 

Our key observation is that we can exclude the elements of the large antichain from the recursion, inserting them back later into each linear extension. Since antichain elements are not constrained by each other, this can be done relatively efficiently. Still, we cannot afford doing the accounting explicitly for every linear extension; instead, we derive a compact formula that captures the overall contribution of the antichain elements.  

Let $B$ denote an antichain of $\cP$ of size $\wid(\cP)$. Let  $C = X \setminus B$ be the set of elements outside $B$, called the \emph{core} of the poset, and denote $k = |C|$. 

For an antichain element $x \in B$, denote by $$D(x) = \{y \in C \mid y \prec x\} \mbox{~~and~~} U(x) = \{y \in C \mid x \prec y\}$$ the sets of core elements below, respectively, above $x$. Clearly, in every linear extension of $\cP$, every element of $D(x)$ appears before $x$, and $x$ appears before every element of $U(x)$.

Let $v_1, \dots, v_k$ denote the elements of the core $C$ indexed according to a linear extension $\sigma = (v_1, \dots, v_k)$ of $\cP[C]$. Let $e(\cP\,|\,\sigma)$ denote the number of linear extensions of $\cP$ that are consistent with $\sigma$, i.e., the ones obtained by inserting the elements of $B$ into $\sigma$. Then, $e(\cP)$ is simply the sum of $e(\cP\,|\,\sigma)$ over all valid linear extensions $\sigma$ of $\cP[C]$. 

Let us fix an ordering $\sigma$, and observe that, unless $D(x)$ is empty, in a linear extension of $\cP$ that extends $\sigma$, every element $x \in B$ must appear after the last element in $D(x)$, which we denote by $\last(x)$. Similarly, unless $U(x)$ is empty, every $x \in B$ must appear before the first element of $U(x)$, denoted $\first(x)$. Note that last and first depend on the ordering $\sigma$. 

In summary, once $\sigma$ is fixed, each element $x \in B$ can be inserted according to an \emph{interval constraint}: after $\last(x)$ and before $\first(x)$, where one or both constraints may be missing. We will count the number of ways $B$ can be inserted into $\sigma$ while observing these constraints. Note that this is for intuition only, as explicitly enumerating each linear extension $\sigma$ of $\cP[C]$ would be inefficient.

\subsection{Making all constraints one-sided }\label{sec:one-sided}

Let $L \subseteq B$ denote the antichain elements that have only a \emph{lower constraint}, i.e., those $x \in B$ where $U(x) = \emptyset$ and $D(x) \neq \emptyset$, and let $T \subseteq B$ denote the antichain elements with \emph{two-sided constraints}, i.e., $x \in B$ where $U(x) \neq \emptyset$ and $D(x) \neq \emptyset$.

A crucial observation is that as we walk along $\sigma$ and attempt to insert elements $x \in B$, those with upper-only constraints (or no constraints at all) are easier to handle, since we can readily compute, once we encounter $\first(x)$ or reach the end of the sequence, the number of possible ways $x$ can be inserted in the sequence of elements already seen. By contrast, lower-only or two-sided constraints are difficult to handle, since they depend on the configuration of yet-unseen elements. 

For this reason, we rewrite all lower constraints as upper constraints with a negative sign, as follows. Denote by $[\cC]$ the truth value of $\cC$, i.e., $[\cC] = 1$ if $\cC$ is true and $[\cC] = 0$ otherwise. Then the equations $[a < x] = 1 - [x < a]$ and $[a<x<b] = [x<b] - [x<a]$ are immediate. Note that here $u < v$ means that $u$ precedes $v$ in an ordering. 

So far we considered inserting one element $x \in B$ into $\sigma$. Such an insertion has to satisfy the interval constraints given by $\first(x)$ and $\last(x)$. An ordering of $X$ is a valid linear extension of $\cP$ (extending $\sigma$) exactly if all elements of $B$ satisfy their interval constraints.

This is captured in the following formula:

\begin{lemma}\label{lem41}
Let $\sigma$ be a linear extension of $\cP[C]$. Then 
$$e(\cP\,|\,\sigma) = \sum_{S \subseteq T} \sum_{R \subseteq L} {(-1)^{|S|+|R|} {N_{\sigma}(S,R)}},$$ where $N_{\sigma}(S,R)$ is the number of orderings of $X$ that order $C$ according to $\sigma$, and satisfy the conditions:
\begin{enumerate}[(i)]
\item $x$ appears before $\last(x)$ for every $x \in S \cup R$,
\item $x$ appears before $\first(x)$ for every $x \in B \setminus S$ where $U(x) \neq \emptyset$.
\end{enumerate}
\end{lemma}

Note that there is no condition imposed for the remaining elements of $B$, i.e., those where $U(x) = \emptyset$ and $x \notin R$.

\begin{proof}
Let $\Pi$ be the set of orderings of $X$ that agree with $\sigma$ on the elements of $C$. For $x \in B$ and $\pi \in \Pi$, let $\cC_{x,\pi}$ denote the condition that $x$ satisfies the constraints imposed by $\first(x)$ and $\last(x)$ in $\pi$. Then, $\pi$ is a valid linear extension of $\cP$ if and only if $\prod_{x \in B}{[\cC_{x,\pi}]} = 1$. Therefore, the total count is 
\begin{align}
e(\cP\,|\,\sigma)  = \sum_{\pi \in \Pi} \prod_{x \in B}{[\cC_{x,\pi}]}\label{eq1}.
\end{align}

We can rewrite $[\cC_{x, \pi}]$, assuming that $\last(x)$ and $\first(x)$ are defined with respect to $\sigma$, as:
\begin{itemize}
\item $[x < \first(x)] - [x < \last(x)]$, if $x$ has a two-sided constraint, i.e., $x \in T$,
\item $1 - [x<\last(x)]$, if $x$ has only a lower constraint, i.e., $x \in L$,
\item $[x < \first(x)]$, if $x$ has only an upper constraint,
\item $1$ if $x$ is unconstrained.
\end{itemize}

For $x \in T \cup L$, write $[\cC_{x, \pi}] = \upalpha_x - \upbeta_x$, where $\upbeta_x = [x < \last(x)]$, and $\upalpha_x = [x < \first(x)]$ if $x \in T$ and $\upalpha_x = 1$ if $x \in L$.

Rewriting the relevant part of the product, we have:
$$\prod_{x \in T \cup L}(\upalpha_x - \upbeta_x) = \sum_{W \subseteq T \cup L} {(-1)^{|W|} \prod_{x \in W}{\upbeta_x} \prod_{x \in (T \cup L)\setminus W}{\upalpha_x}},$$
where $W$ is the set of elements whose negative terms are considered. Split $W$ into $S = W \cap T$ and $R = W \cap L$. Then, the sum splits into a sum over $S \subseteq T$ and a sum over $R \subseteq L$, with $|W| = |S| + |R|$. The factors due to upper-only and unconstrained elements do not depend on the choice of $S$ and $R$ and appear in every term. We thus obtain
\begin{align}
\prod_{x \in B}{[\cC_{x,\pi}]} = \sum_{S \subseteq T} \sum_{R \subseteq L}{(-1)^{|S|+|R|}} {J_{S,R}(\pi)},
\label{eq2}
\end{align} 
where $J_{S,R}(\pi)$ is the product of the following factors:
\begin{itemize}
\item $[x < \last(x)]$ for every $x \in S \cup R$,
\item $[x < \first(x)]$ for every $x \in T \setminus S$ and every upper-only $x$,
\item $1$ for every $x \in L \setminus R$ and every unconstrained $x$.
\end{itemize}

The value $J_{S,R}(\pi)$ is $1$ if and only if $\pi$ satisfies all the conditions listed above. The elements of $T \setminus S$ together with the upper-only elements are exactly the $x \notin S$ with $U(x) \neq \emptyset$. The elements of $L \setminus R$ together with the unconstrained elements are exactly the $x \notin R$ with $U(x) = \emptyset$. So $J_{S,R}(\pi) = 1$ exactly if $\pi$ is one of the orderings counted by $N_{\sigma}(S,R)$, and therefore:
\begin{align}
\sum_{\pi \in \Pi}{J_{S,R}(\pi)} = N_{\sigma}(S,R).
\label{eq3}
\end{align}

Putting everything together, we have 
\begin{align*}
e(\cP\,|\,\sigma)~ & =~ \displaystyle\sum_{\pi \in \Pi} \sum_{S \subseteq T} \sum_{R \subseteq L} {(-1)^{|S|+|R|} \cdot J_{S,R}{(\pi)}} && \triangleright \mbox{~from~\eqref{eq1}, \eqref{eq2}}\\
 & =~ \displaystyle\sum_{S \subseteq T} \sum_{R \subseteq L} {(-1)^{|S|+|R|} \cdot \sum_{\pi \in \Pi} J_{S,R}{(\pi)}}&& \triangleright \mbox{~rearranging}\\
 &=~  \displaystyle\sum_{S \subseteq T} \sum_{R \subseteq L} {(-1)^{|S|+|R|} \cdot N_{\sigma}{(S,R)}}.&& \triangleright \mbox{~from~\eqref{eq3}}\qedhere
\end{align*}
\end{proof}

It remains to compute $N_\sigma(S,R)$.
Note that the definition of this quantity only involves constraints on elements $x \in B$ of the type ``$x$ appears \emph{before} $v$'', for some $v \in C$. In case of such a constraint we say that $x$ is \emph{attached} to $v$. Elements $x \in B$ that have no such constraints are not attached to any element of $C$. 

Walking along $\sigma = (v_1, v_2, \dots, v_k)$, we denote by $a_i$ the number of elements attached to $v_i$. 

\begin{lemma}\label{lem42}
Let $d_0 = 0$, and let $d_i = d_{i-1} + a_i + 1$. Then
$$N_{\sigma}(S,R) = \frac{n!}{d_k!} \prod_{i=1}^k \frac{(d_{i-1} + a_i)!}{d_{i-1}!}. $$
\end{lemma}

Here $d_i = i + a_1 + \cdots + a_i$ is the number of elements appearing up to and including $v_i$; this includes the core elements $v_1, \dots, v_i$ and the antichain elements attached to them. 

\begin{proof}
When considering the core element $v_i$ for $i=1,\dots,k$ (in the sequence given by $\sigma$), we have already placed elements $v_1, \dots, v_{i-1}$, and all elements attached to them, forming a sequence of length $d_{i-1}$. Before appending $v_i$, we need to insert $a_i$ elements attached to $v_i$ into the preceding sequence: the first element has $d_{i-1}+1$ possible positions, the next $d_{i-1}+2$, and so on. This gives, in total, $(d_{i-1}+a_i)!/d_{i-1}!$ different ways. Finally, we append $v_i$. At the very end we insert the $n-d_k$ free elements arbitrarily, in $n!/d_k!$ different ways. 

Note that every ordering counted by $N_{\sigma}(S,R)$ is considered exactly once. Also observe that elements of $B$ are inserted in every possible relative order (which is correct, since these form an antichain and thus there are no precedence constraints between them).
\end{proof}

\subsection{The algorithm} \label{sec:large-alg}
We have derived a formula for $e(\cP\,|\,\sigma)$. It remains to show how to efficiently sum this formula over all valid $\sigma$. First we show how to compute it for a given ordering $\sigma$, and then extend it to all $\sigma$.  

\paragraph{Computing $e(\cP\,|\,\sigma)$.}

Lemmas~\ref{lem41} and \ref{lem42} give a direct way to compute $e(\cP\,|\,\sigma)$: for every pair $(S,R)$, walk along $\sigma$ as in the proof of Lemma~\ref{lem42}, and add up the results with the signs of Lemma~\ref{lem41}. We can make this more efficient, noting that for a fixed $S$, the walks for the different sets $R$ differ only in which elements of $L$ they attach in each round, and by Lemma~\ref{lem42} the rest of the walk depends only on its current length. So a single walk suffices, maintaining a vector $V$ with entries $V[0],\dots,V[n]$: after each round, $V[d]$ is the signed number of ways to have built a sequence of length $d$ so far, added up over all choices of $R$ made so far. 

Consider round $i$ when we place $v=v_i$, after having placed the set $I=\{v_1,\dots,v_{i-1}\}$.  
Let 
\begin{itemize}
  \item $a(I,v)$ be the number of $x\in S$ with $\last(x)=v$, plus the number of $x\notin S$ with $U(x)\ne\emptyset$ and $\first(x)=v$; these elements are attached to $v$ regardless of the choice of $R$;
  \item $c(I,v)$ be the number of $x\in L$ with $\last(x)=v$; such an element is attached to $v$ if $x\in R$, and free otherwise.
\end{itemize}

\begin{algorithm}[h]
\caption{Computing $e(\cP \mid \sigma)$ for a given $\sigma$}
\label{alg_c}
\begin{algorithmic}[1]
\Function{Step}{$V, I, v, S$} \Comment{one round: place $v$ after the set $I$}
  \State $a\gets a(I,v)$; \ $c\gets c(I,v)$
  \State $V'\gets(0,\dots,0)$
  \For{$d=0,\dots,n$}
    \For{$\ell=0,\dots,c$} \Comment{$\ell$ of the $c$ elements are in $R$}
    \If{$d+a+\ell+1\le n$}
      \State $\displaystyle V'[d+a+\ell+1]\mathrel{+}=(-1)^{\ell}\binom{c}{\ell}\frac{(d+a+\ell)!}{d!}\;V[d]$
      \EndIf
    \EndFor
  \EndFor
  \State \Return $V'$
\EndFunction
\Statex
\Function{Walk}{$\sigma, S$} \Comment{$\sigma=(v_1,\dots,v_k)$}
  \State $V\gets(1,0,\dots,0)$; \ $I\gets\emptyset$
  \For{$i=1,\dots,k$}
    \State $V\gets\textproc{Step}(V,I,v_i,S)$; \ $I\gets I\cup\{v_i\}$
  \EndFor
  \State \Return $\sum_{d=0}^n V[d]\cdot n!/d!$ \Comment{insert the $n-d$ free elements}
\EndFunction
\end{algorithmic}
\end{algorithm}

\textsc{Step} executes round~$i$ from the proof of Lemma~\ref{lem42}: it inserts the $a+\ell$ attached elements into a sequence of length $d$, in $(d+a+\ell)!/d!$ ways, and then appends $v$ itself. The binomial coefficient counts the choices of which $\ell$ of the $c$ elements are in $R$; each of them contributes the sign $(-1)$, and all these choices lead to the same length. 

\begin{lemma}\label{lem33}
For every core order $\sigma$ and every $S\subseteq T$,
\[
  \textsc{Walk}(\sigma,S)=\sum_{R\subseteq L}(-1)^{|R|}N_\sigma(S,R).
\]
Consequently, $e(\cP\,|\,\sigma)=\sum_{S\subseteq T}(-1)^{|S|}\,\textsc{Walk}(\sigma,S)$, and $e(\cP)$ is the sum of these values over all linear extensions $\sigma$ of $\cP[C]$.
\end{lemma}

\begin{proof}
Choosing $R\subseteq L$ is the same as choosing, in each round $i$, which of the elements $x\in L$ with $\last(x)=v_i$ belong to $R$. For one such choice, round $i$ of the walk in the proof of Lemma~\ref{lem42} goes from length $d_{i-1}$ to $d_i=d_{i-1}+a+\ell+1$ in $(d_{i-1}+a+\ell)!/d_{i-1}!$ ways, where $a=a(I,v_i)$ and $\ell$ is the number of chosen elements. This is one term of \textsc{Step}, with the sign $(-1)^{\ell}$. So after the last round, $V[d]$ is the sum of $(-1)^{|R|}$ times the product in Lemma~\ref{lem42}, over all $R$ with $d_k=d$. Finally, we multiply by $n!/d!$, completing the formula of Lemma~\ref{lem42}. 
\end{proof}

\paragraph{Computing $\sum{e(\cP\,|\,\sigma)}$ over all valid $\sigma$.}

In \textsc{Step}, the set of elements attached to $v$ only depends on $I$ and $v$, not on the order within $I$. So all orders that reach the same ideal $I$ of $\cP[C]$ can share one vector, just as all terms with the same length share one entry. With exponential memory, we could implement the usual dynamic program over the ideals of $\cP[C]$: 
\[
  \mathit{V}[\emptyset]=(1,0,\dots,0),\qquad \mathit{V}[I\cup\{v\}]\mathrel{+}=\textsc{Step}({V}[I],I,v,S).
\]
Since we can only afford polynomial space, we instead split every ordering of the core in the middle, similarly to Algorithm~\ref{alg_a}. We thus develop a new subroutine, \textsc{Propagate}, described as Algorithm~\ref{alg_d}, which simulates running \textsc{Walk} for all orderings $\sigma$, from a starting ideal $A$ (initially the empty set) to the final ideal $C$ itself, through some intermediate ideal $K$.

\begin{algorithm}[h]
\caption{Computing $e(\cP \mid \sigma)$ for all valid $\sigma$}
\label{alg_d}
\begin{algorithmic}[1]
\Function{Propagate}{$V, A, C, S$} \Comment{$A\subseteq C$ ideals of $\cP[C]$}
  \State $r\gets|C\setminus A|$
  \If{$r=0$}
    \State \Return $V$
  \EndIf
  \If{$r=1$}
    \State \Return $\textproc{Step}(V,A,u,S)$ \Comment{where $C\setminus A=\{u\}$}
  \EndIf
  \State $Y\gets(0,\dots,0)$
  \For{each ideal $K$ of $\cP[C]$ with $A\subseteq K\subseteq C$ and $|K\setminus A|=\lfloor r/2\rfloor$}
    \State $Z\gets\textproc{Propagate}(V,A,K,S)$ \Comment{first half: from $A$ to $K$}
    \State $Y\gets Y+\textproc{Propagate}(Z,K,C,S)$ \Comment{second half: from $K$ to $C$}
  \EndFor
  \State \Return $Y$
\EndFunction
\end{algorithmic}
\end{algorithm}

The candidates $K$ are listed from a minimum chain partition $C_1,\dots,C_m$ of $\cP[C]$, computed once at the start (see Algorithm~\ref{alg_e}). For each ideal $K\supseteq A$, the set $(K \setminus A) \cap C_i$ is a prefix of the chain $C_i\setminus A$, so it is determined by the vector $\big(|(K\setminus A)\cap C_1|,\dots,|(K\setminus A)\cap C_m|\big)$. We list all vectors $(j_1,\dots,j_m)$ with $0\le j_i\le|C_i\setminus A|$ and $j_1+\dots+j_m=\lfloor r/2\rfloor$, one at a time, and keep those that yield an ideal.

\begin{lemma}\label{lem34}
$\textsc{Propagate}(V,A,C,S)$ is the sum, over all linear extensions $(u_1,\dots,u_r)$ of $\cP[C\setminus A]$, of the vector obtained from $V$ by applying $\textsc{Step}(\cdot,A\cup\{u_1,\dots,u_{j-1}\},u_j,S)$ for $j=1,\dots,r$.
\end{lemma}

\begin{proof}
By induction on $r$; for $r\le 1$ it is immediate. Every linear extension of $\cP[C\setminus A]$ passes through exactly one ideal $K$ of the middle size, namely $A$ together with its first $\lfloor r/2\rfloor$ elements. It splits there into a linear extension of $\cP[K\setminus A]$ and one of $\cP[C\setminus K]$. 
Since \textsc{Step} is linear in $V$, adding up the vectors over all first halves, and then running the sum through all second halves, gives the same result as running $V$ through every whole order.
\end{proof}

\paragraph{Putting everything together.} 

We are ready to describe the full algorithm. 
\begin{algorithm}[h]
\caption{Main algorithm for counting linear extensions}
\label{alg_e}
\begin{algorithmic}[1]
\Function{Count}{$\cP$}
  \State $B\gets$ a maximum antichain of $\cP$; \ $C\gets X\setminus B$ 
  \State compute $D(x)$ and $U(x)$ for all $x\in B$, and the sets $T$ and $L$
    \Statex \hfill{\small$\triangleright$ $T$ is used in the loop below; $D$, $U$, $L$ inside \textsc{Step}, via $a(I,v)$ and $c(I,v)$}
  \State compute a minimum chain partition of $\cP[C]$ \Comment{\small used by \textsc{Propagate} to list its candidates}
  \State $\mathit{total}\gets 0$
  \For{each $S\subseteq T$}
    \State $Y\gets\textproc{Propagate}((1,0,\dots,0),\emptyset,C,S)$
    \State $\mathit{total}\gets\mathit{total}+(-1)^{|S|}\sum_{d=0}^n Y[d]\cdot n!/d!$
  \EndFor
  \State \Return $\mathit{total}$
\EndFunction
\end{algorithmic}
\end{algorithm}

Note that \textsc{Count} does not call itself; instead, it calls \textsc{Propagate}, initially going from the empty set to the entire core $C$; \textsc{Propagate} then calls itself recursively with the appropriate balanced splits. 
Correctness follows from Lemmas~\ref{lem33} and~\ref{lem34}. 
In particular, the formula of Lemma~\ref{lem41} sums over the core orders~$\sigma$ and over the sets $S\subseteq T$ and $R\subseteq L$, and \textsc{Count} handles these three sums differently:
\begin{itemize}
  \item the core orders $\sigma$ are summed implicitly, by \textsc{Propagate},
  \item the sets $R$, i.e., lower-only elements attached to $\last{(x)}$, are summed inside \textsc{Step}, by the loop over $\ell$;
  \item the sets $S$, i.e., two-sided elements attached to $\last{(x)}$ instead of $\first{(x)}$, form an explicit outer loop, which yields a factor $2^{|T|}$.
\end{itemize}
The remaining elements of $B$ involve no choice: an upper-only element is always attached to $\first{(x)}$ and is accounted for in $a(I,v)$, and an unconstrained element is always free and is counted through the final factor $n!/d!$.

The recursion of \textsc{Propagate} has depth $\fO(\log k)$, and each level holds a constant number of vectors with $n+1$ entries of $\fO(n\log n)$ bits, so \textsc{Count} uses polynomial space. 

\section{Running time analysis}\label{sec5}

The procedure \textsc{Count} calls \textsc{Propagate}, on the top level subproblem $(\emptyset,C)$, once for each $S\subseteq T$; \textsc{Propagate} then calls itself recursively. All other work, such as listing a candidate vector, testing whether it is an ideal, or running \textsc{Step}, takes polynomial time per call or per listed vector.

Let $N(A,A')$ denote the total number of candidate vectors that \textsc{Propagate} lists when it is called on the subproblem\footnote{A \emph{subproblem} is a pair $(A,A')$ of ideals of $\cP[C]$, where $A \subseteq A' \subseteq C$. Such a subproblem is solved by \textsc{Propagate}$(V, A, A', S)$.} $(A,A')$, over all levels of its recursion. This number does not depend on $V$ or $S$.

The procedure \textsc{Propagate} lists some vectors at the top level, and recurses on the subproblems $(A,K)$ and $(K,A')$ for each listed $K$ that is an ideal. So $N(A,A')=0$ if $|A'\setminus A|\le 1$, and otherwise
  \[
    N(A,A')=(\text{number of vectors listed at the top level})+\sum_{K}\big(N(A,K)+N(K,A')\big).
  \]
The call itself takes time $\fO^*(N(A,A')+1)$, and therefore \textsc{Count} takes time $\fO^*\big(2^{|T|}(N(\emptyset,C)+1)\big)$. In \S\,\ref{sec41} we bound $N(\emptyset,C)$, and in \S\,\ref{sec42} we complete the proof of Theorem~\ref{thm1}.

\subsection{\textsc{Propagate}}\label{sec41}

If the core is an antichain, every subset of the right size is a candidate, and $N(\emptyset,C)$ is $4^k$ up to a polynomial factor, as in the baseline algorithm. Chains in the core reduce this: we will see that only the first element of a chain contributes a factor~$4$, and every further element a factor of at most $8/3$.

A relation $x\prec y$ in $\cP[C]$ rules out every candidate that contains $y$ but not $x$. At the top level, this saves a factor $4/3$, since the pair has $3$ possible prefixes instead of $4$. The pair $x,y$ again leads to savings at every lower level where they fall into the same subproblem, and we will account for the savings at all levels. 

\paragraph{Chain weights.} Let $C_1,\dots,C_m$ be the minimum chain partition of $\cP[C]$ from which \textsc{Propagate} lists its candidates, with lengths $\ell_1,\dots,\ell_m$. 
Consider a poset $Q$ consisting of the same chains, but with no relations between different chains. In every subproblem, \textsc{Propagate} lists the same candidate vectors for $\cP[C]$ as for $Q$, and it recurses only on those that are ideals of $\cP[C]$. Every ideal of $\cP[C]$ is an ideal of $Q$, so every subproblem that arises for $\cP[C]$ also arises for $Q$.
Hence, $N(\emptyset,C)$ is \emph{at most} the number of candidate vectors for $Q$, which 
depends only on $\ell_1,\dots,\ell_m$; we denote it by $N(\ell_1,\dots,\ell_m)$.

We assign each chain a weight. This is only for the purpose of the analysis; the algorithm does not compute them. The weights bound how much each chain contributes to the number of candidates, over all levels of the recursion together. In the antichain case, the count $4^k$ can be expressed as a factor 4 for each element. In case of a chain, the contribution should be less than 4 per element. 

We wish to set $\mu(\ell)$ for each chain length $\ell$, so that $N(\ell_1,\dots,\ell_m)$ is at most a polynomial times $\Phi=\mu(\ell_1)\cdots\mu(\ell_m)$. The induction in the following lemma goes through if the weights satisfy
\[
  \sum_{j=0}^{\ell}\mu(j)\,4^{-j}=2^{-\ell}\mu(\ell)\qquad\text{for all }\ell\ge 0. \tag{$\star$}
\]
Roughly, $(\star)$ says that the first halves, added up over all candidates, weigh no more than the whole subproblem. Starting from $\mu(0)=1$, subtracting $(\star)$ for $\ell-1$ from $(\star)$ for $\ell$ determines each weight from the previous one:
\[
  \mu(\ell)=\frac{2\mu(\ell-1)}{1-2^{-\ell}},\qquad \mu(1)=4,\quad \mu(2)=\tfrac{32}{3},\quad \mu(3)=\tfrac{512}{21}.
\]
Conversely, weights defined by this rule satisfy $(\star)$, by induction on $\ell$. So a single element weighs 4, as in the baseline, while a chain of two elements weighs $32/3 \approx 10.67$ instead of $4^2=16$.

\begin{lemma}\label{lem10}
For chains of lengths $\ell_1,\dots,\ell_m$ with $r=\ell_1+\dots+\ell_m\ge 1$ elements in total, $N(\ell_1,\dots,\ell_m)\le r^3\cdot\Phi$.
\end{lemma}

The idea of the proof is to count the top-level candidates, and to apply induction to their two halves. Condition~$(\star)$ guarantees that the weights of the halves, summed over all candidates, can be factored over the chains.

\begin{proof}

By induction on $r$; for $r\le 2$ the claim is immediate. Let $r\ge 3$. For an integer $t$, let
\[
  J_t=\big\{(j_1,\dots,j_m)\in\mathbb{Z}^m \;:\; 0\le j_i\le\ell_i \text{ for all } i, \text{ and } j_1+\dots+j_m=t\big\}.
\]
At the top level, the listed vectors are the elements of $J_{\lfloor r/2\rfloor}$. There are at most $(\ell_1+1)\cdots(\ell_m+1)\le\Phi$ of them, since $\ell+1\le\mu(\ell)$. Each $(j_1,\dots,j_m)\in J_{\lfloor r/2\rfloor}$ gives two subproblems, with chain lengths $(j_1,\dots,j_m)$ and $(\ell_1-j_1,\dots,\ell_m-j_m)$, each with at most $\lceil r/2\rceil$ elements. By the recurrence for $N$ and by induction,
\[
  N(\ell_1,\dots,\ell_m)\;\le\;\Phi+\lceil r/2\rceil^3\sum_{(j_1,\dots,j_m)\in J_{\lfloor r/2\rfloor}}\Big(\prod_{i=1}^m\mu(j_i)+\prod_{i=1}^m\mu(\ell_i-j_i)\Big).
\]
Since $(j_1,\dots,j_m)\mapsto(\ell_1-j_1,\dots,\ell_m-j_m)$ is a bijection from $J_{\lfloor r/2\rfloor}$ to $J_{\lceil r/2\rceil}$, the right side equals $\Phi+\lceil r/2\rceil^3\big(S_{\lfloor r/2\rfloor}+S_{\lceil r/2\rceil}\big)$, where
\[
  S_t=\sum_{(j_1,\dots,j_m)\in J_t}\ \prod_{i=1}^m\mu(j_i).
\]
For $(j_1,\dots,j_m)\in J_t$ we have $\prod_i\mu(j_i)=4^t\prod_i\mu(j_i)\,4^{-j_i}$, since $j_1+\dots+j_m=t$. Summing over all vectors with $0\le j_i\le\ell_i$, without the condition on the total, only adds nonnegative terms, and this larger sum factors over the chains. With~$(\star)$ we get
\begin{align*}
  S_t\;&\le\;4^t\sum_{j_1=0}^{\ell_1}\cdots\sum_{j_m=0}^{\ell_m}\ \prod_{i=1}^m\mu(j_i)\,4^{-j_i}
  \;=\;4^t\prod_{i=1}^m\ \sum_{j=0}^{\ell_i}\mu(j)\,4^{-j}\\
  &=\;4^t\prod_{i=1}^m 2^{-\ell_i}\mu(\ell_i)\;=\;4^t\,2^{-r}\,\Phi .
\end{align*}
Hence $S_{\lfloor r/2\rfloor}\le\Phi$ and $S_{\lceil r/2\rceil}\le 2\Phi$, and $N(\ell_1,\dots,\ell_m)\le(1+3\lceil r/2\rceil^3)\,\Phi\le r^3\,\Phi$, where the last step holds for $r\ge 3$.
\end{proof}

\paragraph{Long chains.} By the recurrence for $\mu$, the ratio $\mu(\ell)/\mu(\ell-1)=2/(1-2^{-\ell})$ is at most $8/3$ for $\ell\ge2$. So the first element of a chain weighs $4$, and every further element contributes a factor of at most $8/3$:
\[
  \mu(\ell)\le 4\cdot(8/3)^{\ell-1}\le 4^{\ell}\qquad\text{for all }\ell\ge1. \tag{$\star\star$}
\]

\subsection{Overall running time}\label{sec42}

\begin{lemma}\label{lem13}
Let $\ell_1,\dots,\ell_m$ be the chain lengths of a minimum chain partition of $\cP[C]$, and let $\Phi=\mu(\ell_1)\cdots\mu(\ell_m)$. Then $2^{n-k}\cdot\Phi\le(2\sqrt2)^n$.
\end{lemma}

\begin{proof}
Let $w=n-k=|B|$. Then $m\le w$, since an antichain of $\cP[C]$ is also an antichain of $\cP$. If $k\le w$, then $\Phi\le4^k$ by~$(\star\star)$, and $2^{n-k}\Phi\le2^{n+k}\le2^{3n/2}$. If $k\ge w$, then $\Phi\le4^m(8/3)^{k-m}\le4^w(8/3)^{k-w}$ by~$(\star\star)$, as $m\le w$ and $8/3<4$. Using $n=2w+(k-w)$, we get
\[
  2^{n-k}\,\Phi\;\le\;2^w\cdot4^w\Big(\frac83\Big)^{k-w}\;=\;8^w\Big(\frac83\Big)^{k-w}\;\le\;8^w\big(2\sqrt2\big)^{k-w}\;=\;\big(2\sqrt2\big)^{n}. \qedhere
\]
\end{proof}

\begin{proof}[Proof of Theorem~\ref{thm1}]
Algorithm~\ref{alg_e} is correct and uses polynomial space, as shown in \S\,\ref{sec:large-alg}. It runs in time $\fO^*\big(2^{|T|}(N(\emptyset,C)+1)\big)$. Here $2^{|T|}\le2^{|B|}=2^{n-k}$, and $N(\emptyset,C)\le k^3\Phi$ by the reduction to $Q$ in \S\,\ref{sec41} and Lemma~\ref{lem10}. By Lemma~\ref{lem13}, the running time is $\fO^*\big((2\sqrt2)^n\big)$.
\end{proof}

\section{Algorithm for jump number}\label{sec:jump}

In this section we prove Theorem~\ref{thm:jump}. For a linear extension $\pi$ of $\cP$, let $\mathrm{bump}(\pi)$ be its number of bumps, and let $\mathrm{bump}(\cP)$ be the maximum of $\mathrm{bump}(\pi)$ over all linear extensions $\pi$ of $\cP$. Then $\mathrm{jump}(\cP)=n-1-\mathrm{bump}(\cP)$, and we compute $\mathrm{bump}(\cP)$.

The algorithm is a variant of \textsc{Count} (Algorithm~\ref{alg_e}). We keep the notation of \S\,\ref{sec4}: $B$ is a maximum antichain of $\cP$, $C=X\setminus B$ is the core, $k=|C|$, and for $x\in B$ the sets $D(x)$ and $U(x)$ contain the core elements below and above~$x$. Again, $L$ is the set of lower-only and $T$ the set of two-sided elements of $B$. 
Compared to \textsc{Count}, three things change.
\begin{itemize}
  \item The inclusion-exclusion of \S\,\ref{sec:one-sided} is not needed: an element of $B$ can form a bump only in one of two positions, which the walk along an ordering of the core can detect locally (\S\,\ref{sec:jump-fixed}).
  \item In the outer loop over $S\subseteq T$ we now choose, for every two-sided element, which of its two positions it may use.
  \item We maintain a vector indexed by the last placed core element instead of the length of the sequence, with sums and products replaced by maxima and sums (\S\,\ref{sec:jump-alg}).
\end{itemize}
The running-time analysis of \S\,\ref{sec5} carries over essentially unchanged (\S\,\ref{sec:jump-time}).

\subsection{Bumps for a fixed ordering}\label{sec:jump-fixed}

Fix a linear extension $\sigma=(v_1,\dots,v_k)$ of $\cP[C]$, with $\mathrm{last}(x)$ and $\mathrm{first}(x)$ defined with respect to $\sigma$ as in \S\,\ref{sec4}. A linear extension $\pi$ of $\cP$ that orders $C$ as $\sigma$ places the elements of $B$ into the $k+1$ \emph{gaps} of $\sigma$: before $v_1$, between $v_i$ and $v_{i+1}$ for $1\le i<k$, and after~$v_k$.

A bump cannot occur between two elements of $B$, as $B$ is an antichain. If $x\in B$ is immediately preceded in $\pi$ by a core element $y$ with $y\prec x$, then $y=\mathrm{last}(x)$, because all elements of $D(x)$ precede~$x$. Symmetrically, if $x$ is immediately followed by a core element $z$ with $x\prec z$, then $z=\mathrm{first}(x)$. So $x$ can form a bump only in two positions: as the first element of the gap after $\mathrm{last}(x)$, which we call its \emph{low position}, and as the last element of the gap before $\mathrm{first}(x)$, its \emph{high position}. Lower-only elements have only a low position, upper-only elements only a high position, and unconstrained elements never form a bump. Two-sided elements have both positions, and the two positions of such an $x$ are in the same gap exactly when $\mathrm{last}(x)$ and $\mathrm{first}(x)$ are consecutive in~$\sigma$; then $x$ can form two bumps, if it is the only element placed in this gap. 

\paragraph{Two-sided elements.}
As we walk along $\sigma$, we can detect $\mathrm{last}(x)$ and $\mathrm{first}(x)$ as we reach them, so one-sided elements cause no difficulty. As for \#LE, the main issue is handling two-sided elements: whether such an element uses its low or its high position (assuming they fall into different gaps) affects two gaps, which may be far apart in~$\sigma$. We fix these choices in an outer loop: for each $S\subseteq T$, the elements of $S$ may use only their low position, and the elements of $T\setminus S$ only their high position.
An element whose two positions are in the same gap may use both positions, regardless of~$S$.

\paragraph{The value of a gap.}
Consider the gap between two consecutive core elements $p$ and $v$ of $\sigma$, and let $I$ be the set of core elements up to and including~$p$. To treat the first and the last gap in the same way, we let $p=\bot$ if $I=\emptyset$, and $v=\top$ if $I=C$, with $[\bot\prec v]=[p\prec\top]=0$. As in \S\,\ref{sec:large-alg}, the relevant facts depend only on $I$, $p$ and~$v$: $p=\mathrm{last}(x)$ exactly when $p\in D(x)\subseteq I$, and $v=\mathrm{first}(x)$ exactly when $v\in U(x)$ and $U(x)\cap I=\emptyset$. For $S\subseteq T$, the \emph{value} of the gap is
\[
  g_S(I,p,v)=\max\begin{cases}
    [p\prec v],\\[2pt]
    [\exists x\in L\cup S:\ p=\mathrm{last}(x)]+[\exists x\in B\setminus(L\cup S):\ v=\mathrm{first}(x)],\\[2pt]
    2\cdot[\exists x\in T:\ p=\mathrm{last}(x)\text{ and }v=\mathrm{first}(x)].
  \end{cases}
\]

The three terms correspond to leaving the gap empty; placing an element in its low position right after $p$ and/or an element in its high position right before $v$; and placing a single two-sided element between $p$ and~$v$. These values are an upper bound on the number of bumps for a gap; the next lemma shows that the values can actually be realized.

\begin{lemma}\label{lem:jump-fixed}
Let $\sigma=(v_1,\dots,v_k)$ be a linear extension of $\cP[C]$, let $I_i=\{v_1,\dots,v_i\}$ for $0\le i\le k$, and let $v_0=\bot$ and $v_{k+1}=\top$. The maximum of $\mathrm{bump}(\pi)$ over all linear extensions $\pi$ of $\cP$ that order $C$ as $\sigma$ equals
\[
  \max_{S\subseteq T}\ \sum_{i=0}^{k} g_S(I_i,v_i,v_{i+1}).
\]
\end{lemma}

\begin{proof}
Let $\pi$ be a linear extension of $\cP$ that orders $C$ as $\sigma$, and let $S$ be the set of $x\in T$ that immediately follow $\mathrm{last}(x)$ in~$\pi$. We count the bumps of $\pi$ gap by gap: to the gap between $p=v_i$ and $v=v_{i+1}$ we assign the bumps among $p$, the elements of the gap, and~$v$. If the gap is empty, it has at most $[p\prec v]$ bumps. Otherwise, let $x$ and $x'$ be its first and its last element. If $(p,x)$ is a bump, then $p=\mathrm{last}(x)$, and $x\in L\cup S$ by the choice of~$S$. If $(x',v)$ is a bump, then $v=\mathrm{first}(x')$. If moreover $x'\in L\cup S$, then $x'\in S$ immediately follows $\mathrm{last}(x')$; so $x'=x$ is alone in the gap, $p=\mathrm{last}(x)$ and $v=\mathrm{first}(x)$, and the third term equals~$2$. In all cases, the gap has at most $g_S(I_i,v_i,v_{i+1})$ bumps.

Conversely, fix $S\subseteq T$; we construct $\pi$ with at least $\sum_i g_S(I_i,v_i,v_{i+1})$ bumps. Assign each $x\in L\cup S$ to the gap of its low position, and each $x\in B\setminus(L\cup S)$ with $U(x)\neq\emptyset$ to the gap of its high position. Then every element is assigned to at most one gap, and every element whose existence is required by a term of $g_S(I_i,v_i,v_{i+1})$ is assigned to the gap between $v_i$ and $v_{i+1}$; for the third term, both positions of the element lie in this gap. In every gap, realize a term attaining the maximum: leave the gap empty, or place an element in its low position and/or an element in its high position, or place a single two-sided element. These placements respect the constraints of the placed elements, since $\mathrm{last}(x)$ precedes $\mathrm{first}(x)$ in $\sigma$ for every two-sided~$x$. Finally, insert every remaining $x\in B$ immediately after $\mathrm{last}(x)$, or at the very beginning if $D(x)=\emptyset$. This respects the constraints of~$x$, and it does not decrease the number of bumps: an insertion at the beginning breaks no pair, and an insertion after $\mathrm{last}(x)$ can only break the pair formed by $\mathrm{last}(x)$ and its successor, while it creates the bump $(\mathrm{last}(x),x)$.
\end{proof}

\subsection{The algorithm}\label{sec:jump-alg}

By Lemma~\ref{lem:jump-fixed}, $\mathrm{bump}(\cP)$ is the maximum, over all $S\subseteq T$ and all linear extensions $\sigma$ of $\cP[C]$, of the total value of the gaps of~$\sigma$. For a fixed $S$, we compute the maximum over $\sigma$ in the same way as the sum over $\sigma$ in \S\,\ref{sec:large-alg}, with two changes.

First, the vector carried along the walk records the last placed core element instead of the length of the sequence. It is indexed by $C\cup\{\bot\}$, and after placing a set $I$, the entry $V[p]$ is the largest total value of the gaps passed so far, over all linear extensions of $\cP[I]$ that end with~$p$, or $-\infty$ if there is no such order. Placing the next element $v$ passes the gap between the last element $p$ and~$v$. This is done by \textsc{BumpStep} in Algorithm~\ref{alg:jump}, which replaces \textsc{Step}.

Second, as we maximize instead of counting, sums of vectors become entrywise maxima. Let \textsc{MaxPropagate} be \textsc{Propagate} (Algorithm~\ref{alg_d}) with \textsc{BumpStep} in place of \textsc{Step}, with $Y$ initialized to $(-\infty,\dots,-\infty)$, and with the sum $Y+\textsc{Propagate}(Z,K,C,S)$ replaced by the entrywise maximum of $Y$ and $\textsc{MaxPropagate}(Z,K,C,S)$. Lemma~\ref{lem34} holds for \textsc{MaxPropagate}, with ``sum'' replaced by ``maximum'' and \textsc{Step} by \textsc{BumpStep}. The proof is the same, since \textsc{BumpStep} commutes with entrywise maxima, just as \textsc{Step} commutes with sums.

\begin{algorithm}[t]
\caption{Computing the jump number}\label{alg:jump}
\begin{algorithmic}[1]
\Function{BumpStep}{$V,I,v,S$}\Comment{pass the gap before $v$, then $v$ is last}
  \State $V'\gets(-\infty,\dots,-\infty)$
  \State $V'[v]\gets\max_{p\in I\cup\{\bot\}}\big(V[p]+g_S(I,p,v)\big)$
  \State \Return $V'$
\EndFunction
\Statex
\Function{JumpNumber}{$\cP$}
  \State $B\gets$ a maximum antichain of $\cP$; \ $C\gets X\setminus B$
  \State compute $D(x)$ and $U(x)$ for all $x\in B$, and the sets $T$ and $L$
  \State compute a minimum chain partition of $\cP[C]$\Comment{used by \textsc{MaxPropagate}}
  \State $\mathit{best}\gets-\infty$
  \For{each $S\subseteq T$}
    \State $Y\gets\textproc{MaxPropagate}(e_\bot,\emptyset,C,S)$\Comment{$e_\bot[\bot]=0$, other entries $-\infty$}
    \State $\mathit{best}\gets\max\big(\mathit{best},\ \max_{p}\,(Y[p]+g_S(C,p,\top))\big)$\Comment{add the last gap}
  \EndFor
  \State \Return $n-1-\mathit{best}$
\EndFunction
\end{algorithmic}
\end{algorithm}

\paragraph{Correctness and space.}
For each $S$, the entry $Y[p]$ is the largest total value of all gaps but the last, over all linear extensions of $\cP[C]$ that end with~$p$, by Lemma~\ref{lem34} for \textsc{MaxPropagate}. Adding the value of the last gap and maximizing over $p$ and $S$ gives $\mathrm{bump}(\cP)$, by Lemma~\ref{lem:jump-fixed}. The space is polynomial, as for \textsc{Count}: the recursion has depth $\mathcal{O}(\log k)$, and each level stores a constant number of vectors with $k+1$ entries from $\{-\infty,0,1,\dots,n\}$.

A linear extension with $\mathrm{jump}(\cP)$ jumps can be found within the same bounds by the usual self-reduction: for an optimal $S$, we recompute the top level of \textsc{MaxPropagate} to find a middle ideal $K$ and a last element of $K$ through which an optimal order passes, recurse on the two halves, and finally place $B$ as in the proof of Lemma~\ref{lem:jump-fixed}.

\subsection{Running time}\label{sec:jump-time}

The procedures call each other similarly as for \#LE: \textsc{JumpNumber} calls \textsc{MaxPropagate} on the subproblem $(\emptyset,C)$ once for each $S\subseteq T$, and \textsc{MaxPropagate} calls itself and \textsc{BumpStep}. Each listed candidate and each call of \textsc{BumpStep} takes polynomial time. So, by the recurrences of \S\,\ref{sec5}, \textsc{JumpNumber} runs in time $\mathcal{O}^*\big(2^{|T|}(N(\emptyset,C)+1)\big)$, the same bound as \textsc{Count}.

\begin{proof}[Proof of Theorem~\ref{thm:jump}]
Correctness and polynomial space were shown in \S\,\ref{sec:jump-alg}. The running time is $\mathcal{O}^*\big(2^{|T|}(N(\emptyset,C)+1)\big)$, which is $\mathcal{O}^*\big((2\sqrt2)^n\big)$ by Lemmas~\ref{lem10} and~\ref{lem13}, exactly as in the proof of Theorem~\ref{thm1}.
\end{proof}

\section{Conclusions}\label{sec:concl}
We gave deterministic polynomial-space algorithms for counting linear extensions and for the jump number problem, running in time $\fO^*((2\sqrt{2})^{n}) \subseteq \fO(2.829^n)$. We expect the technique to be applicable to other poset problems, and perhaps also to a broader class of permutation problems. Extending the result to TSP seems difficult; we note that solving TSP in polynomial space and time $\fO((4-\varepsilon)^n)$ remains open.

Further improving the running time for \#LE remains an interesting direction. 
We note that with AI assistance, we were able to improve the running time of our approach to $\fO(2.7693^n)$, still in polynomial space. 

At a high level, the analysis of the $\fO^*((2\sqrt{2})^n)$-algorithm is tight only when the core is close to an antichain and almost all elements of $B$ have two-sided constraints. These two situations are in tension, since every two-sided element creates relations in the core, pushing it further from being an antichain. An improvement arises from combining two ideas. First, the choice made for a two-sided element only matters while the walk is between its last predecessor and its first successor in the core, so instead of fixing all choices in advance, they can be made on the fly, which is cheap when these windows are short. Second, by Menger's theorem, either there are many disjoint configurations $a\prec x\prec c$ with $x\in B$ and $a,c\in C$, in which case the windows can be made short, or a few core elements, called hubs, cover the relations of almost all two-sided elements, in which case these elements are grouped by their hubs, and each group can be handled at once. The grouping by hubs is similar to techniques used in~\cite{Oka}.  
Since the details are rather complex and the overall improvement is not large, we decided to omit a full description of this improved algorithm.

The improvement does not readily transfer to the jump number problem, so a polynomial-space algorithm faster than $(2\sqrt{2})^n$ remains open. We mention two other long-standing open questions about the jump number problem: (1) parameterized by the number $s$ of jumps, the problem can be solved in time $s^{\fO(s)} \mathrm{poly}(n)$~\cite{McC01}; can the dependence on $s$ be improved to $2^{\fO(s)}$? (2) can the jump number problem be solved in polynomial time for two-dimensional posets~\cite{SS87,Cer03}?

The recently claimed refutation of the exponential-time hypothesis (ETH) involved a \emph{subexponential} algorithm for counting $k$-CNF solutions~\cite{openai}. As we are not aware of a reduction from \#LE to \#$k$SAT with a nearly linear number of variables, these results do not appear directly applicable to our problems.    

\paragraph{AI methodology.}

The results of the paper were obtained with the assistance of the AI tools GPT Astra 6.0 and Claude Opus 5.5, through a process of interactive prompting and refining. The authors laid out the balanced splitting approach and the difficulty of dealing with a large antichain beyond two-dimensional posets, which we handled in past work. The AI tool then generated the majority of the technical work, in particular the core idea of inverting ``after-constraints'', that allowed the inclusion-exclusion formulation of Lemma~\ref{lem41} in order to exclude a large antichain from the recursion. The authors then distilled and refined the ideas towards a more intuitive presentation. The same tools were finally used to verify and reformulate technical arguments; the decisions on whether to incorporate suggestions were made on a case-by-case basis.

\small

\printbibliography
\end{document}